\PassOptionsToPackage{unicode}{hyperref}
\PassOptionsToPackage{hyphens}{url}
\PassOptionsToPackage{dvipsnames,svgnames,x11names}{xcolor}
\documentclass[
]{article}
\usepackage{amsmath,amssymb}
\usepackage{lmodern}
\usepackage{iftex}
\usepackage[T1]{fontenc}
\usepackage[utf8]{inputenc}
\usepackage{textcomp} 

\IfFileExists{upquote.sty}{\usepackage{upquote}}{}
\IfFileExists{microtype.sty}{
  \usepackage[]{microtype}
  \UseMicrotypeSet[protrusion]{basicmath} 
}{}
\makeatletter
\@ifundefined{KOMAClassName}{
  \IfFileExists{parskip.sty}{%
    \usepackage{parskip}
  }{
    \setlength{\parindent}{0pt}
    \setlength{\parskip}{6pt plus 2pt minus 1pt}}
}{
  \KOMAoptions{parskip=half}}
\makeatother
\usepackage{xcolor}
\providecommand{\tightlist}{%
  \setlength{\itemsep}{0pt}\setlength{\parskip}{0pt}}
\NewDocumentCommand\citeproctext{}{}
\NewDocumentCommand\citeproc{mm}{%
  \begingroup\def\citeproctext{#2}\cite{#1}\endgroup}
\makeatletter
 \let\@cite@ofmt\@firstofone
 \def\@biblabel#1{}
 \def\@cite#1#2{{#1\if@tempswa , #2\fi}}
\makeatother
\newlength{\cslhangindent}
\newlength{\csllabelwidth}
\newenvironment{CSLReferences}[2] 
 {\begin{list}{}{%
  \setlength{\itemindent}{0pt}
  \setlength{\leftmargin}{0pt}
  \setlength{\parsep}{0pt}
  \ifodd #1
   \setlength{\leftmargin}{\cslhangindent}
   \setlength{\itemindent}{-1\cslhangindent}
  \fi
  \setlength{\itemsep}{#2\baselineskip}}}
 {\end{list}}
\usepackage{calc}

\ifLuaTeX
\usepackage[bidi=basic]{babel}
\else
\usepackage[bidi=default]{babel}
\fi
\babelprovide[main,import]{american}

\def\languageshorthands#1{}
\ifLuaTeX
  \usepackage{selnolig}  
\fi
\IfFileExists{bookmark.sty}{\usepackage{bookmark}}{\usepackage{hyperref}}
\IfFileExists{xurl.sty}{\usepackage{xurl}}{} 
\hypersetup{
  pdftitle={Improving reproducibility of cheminformatics workflows with
chembl-downloader},
  pdfauthor={Charles Tapley Hoyt},
  pdflang={en-US},
  colorlinks=true,
  linkcolor={Maroon},
  filecolor={Maroon},
  citecolor={Blue},
  urlcolor={Blue},
  pdfcreator={LaTeX via pandoc}}

\title{Improving reproducibility of cheminformatics workflows with
\texttt{chembl-downloader}}

\definecolor{c53baa1}{RGB}{83,186,161}
\definecolor{c202826}{RGB}{32,40,38}
\def \rorglobalscale {0.1}
\newcommand{\rorlogo}{%
\begin{tikzpicture}[y=1cm, x=1cm, yscale=\rorglobalscale,xscale=\rorglobalscale, every node/.append style={scale=\rorglobalscale}, inner sep=0pt, outer sep=0pt]
  \begin{scope}[even odd rule,line join=round,miter limit=2.0,shift={(-0.025, 0.0216)}]
    \path[fill=c53baa1,nonzero rule,line join=round,miter limit=2.0] (1.8164, 3.012) -- (1.4954, 2.5204) -- (1.1742, 3.012) -- (1.8164, 3.012) -- cycle;
    \path[fill=c53baa1,nonzero rule,line join=round,miter limit=2.0] (3.1594, 3.012) -- (2.8385, 2.5204) -- (2.5172, 3.012) -- (3.1594, 3.012) -- cycle;
    \path[fill=c53baa1,nonzero rule,line join=round,miter limit=2.0] (1.1742, 0.0669) -- (1.4954, 0.5588) -- (1.8164, 0.0669) -- (1.1742, 0.0669) -- cycle;
    \path[fill=c53baa1,nonzero rule,line join=round,miter limit=2.0] (2.5172, 0.0669) -- (2.8385, 0.5588) -- (3.1594, 0.0669) -- (2.5172, 0.0669) -- cycle;
    \path[fill=c202826,nonzero rule,line join=round,miter limit=2.0] (3.8505, 1.4364).. controls (3.9643, 1.4576) and (4.0508, 1.5081) .. (4.1098, 1.5878).. controls (4.169, 1.6674) and (4.1984, 1.7642) .. (4.1984, 1.8777).. controls (4.1984, 1.9719) and (4.182, 2.0503) .. (4.1495, 2.1132).. controls (4.1169, 2.1762) and (4.0727, 2.2262) .. (4.0174, 2.2635).. controls (3.9621, 2.3006) and (3.8976, 2.3273) .. (3.824, 2.3432).. controls (3.7505, 2.359) and (3.6727, 2.367) .. (3.5909, 2.367) -- (2.9676, 2.367) -- (2.9676, 1.8688).. controls (2.9625, 1.8833) and (2.9572, 1.8976) .. (2.9514, 1.9119).. controls (2.9083, 2.0164) and (2.848, 2.1056) .. (2.7705, 2.1791).. controls (2.6929, 2.2527) and (2.6014, 2.3093) .. (2.495, 2.3487).. controls (2.3889, 2.3881) and (2.2728, 2.408) .. (2.1468, 2.408).. controls (2.0209, 2.408) and (1.905, 2.3881) .. (1.7986, 2.3487).. controls (1.6925, 2.3093) and (1.6007, 2.2527) .. (1.5232, 2.1791).. controls (1.4539, 2.1132) and (1.3983, 2.0346) .. (1.3565, 1.9436).. controls (1.3504, 2.009) and (1.3351, 2.0656) .. (1.3105, 2.1132).. controls (1.2779, 2.1762) and (1.2338, 2.2262) .. (1.1785, 2.2635).. controls (1.1232, 2.3006) and (1.0586, 2.3273) .. (0.985, 2.3432).. controls (0.9115, 2.359) and (0.8337, 2.367) .. (0.7519, 2.367) -- (0.1289, 2.367) -- (0.1289, 0.7562) -- (0.4837, 0.7562) -- (0.4837, 1.4002) -- (0.6588, 1.4002) -- (0.9956, 0.7562) -- (1.4211, 0.7562) -- (1.0118, 1.4364).. controls (1.1255, 1.4576) and (1.2121, 1.5081) .. (1.2711, 1.5878).. controls (1.2737, 1.5915) and (1.2761, 1.5954) .. (1.2787, 1.5991).. controls (1.2782, 1.5867) and (1.2779, 1.5743) .. (1.2779, 1.5616).. controls (1.2779, 1.4327) and (1.2996, 1.3158) .. (1.3428, 1.2113).. controls (1.3859, 1.1068) and (1.4462, 1.0176) .. (1.5237, 0.944).. controls (1.601, 0.8705) and (1.6928, 0.8139) .. (1.7992, 0.7744).. controls (1.9053, 0.735) and (2.0214, 0.7152) .. (2.1474, 0.7152).. controls (2.2733, 0.7152) and (2.3892, 0.735) .. (2.4956, 0.7744).. controls (2.6016, 0.8139) and (2.6935, 0.8705) .. (2.771, 0.944).. controls (2.8482, 1.0176) and (2.9086, 1.1068) .. (2.952, 1.2113).. controls (2.9578, 1.2253) and (2.9631, 1.2398) .. (2.9681, 1.2544) -- (2.9681, 0.7562) -- (3.3229, 0.7562) -- (3.3229, 1.4002) -- (3.4981, 1.4002) -- (3.8349, 0.7562) -- (4.2603, 0.7562) -- (3.8505, 1.4364) -- cycle(0.9628, 1.7777).. controls (0.9438, 1.7534) and (0.92, 1.7357) .. (0.8911, 1.7243).. controls (0.8623, 1.7129) and (0.83, 1.706) .. (0.7945, 1.7039).. controls (0.7588, 1.7015) and (0.7252, 1.7005) .. (0.6932, 1.7005) -- (0.4839, 1.7005) -- (0.4839, 2.0667) -- (0.716, 2.0667).. controls (0.7477, 2.0667) and (0.7805, 2.0643) .. (0.8139, 2.0598).. controls (0.8472, 2.0553) and (0.8768, 2.0466) .. (0.9025, 2.0336).. controls (0.9282, 2.0206) and (0.9496, 2.0021) .. (0.9663, 1.9778).. controls (0.9829, 1.9534) and (0.9914, 1.9209) .. (0.9914, 1.8799).. controls (0.9914, 1.8362) and (0.9819, 1.8021) .. (0.9628, 1.7777) -- cycle(2.6125, 1.3533).. controls (2.5889, 1.2904) and (2.5553, 1.2359) .. (2.5112, 1.1896).. controls (2.4672, 1.1433) and (2.4146, 1.1073) .. (2.3529, 1.0814).. controls (2.2916, 1.0554) and (2.2228, 1.0427) .. (2.1471, 1.0427).. controls (2.0712, 1.0427) and (2.0026, 1.0557) .. (1.9412, 1.0814).. controls (1.8799, 1.107) and (1.8272, 1.1433) .. (1.783, 1.1896).. controls (1.7391, 1.2359) and (1.7052, 1.2904) .. (1.6817, 1.3533).. controls (1.6581, 1.4163) and (1.6465, 1.4856) .. (1.6465, 1.5616).. controls (1.6465, 1.6359) and (1.6581, 1.705) .. (1.6817, 1.7687).. controls (1.7052, 1.8325) and (1.7388, 1.8873) .. (1.783, 1.9336).. controls (1.8269, 1.9799) and (1.8796, 2.0159) .. (1.9412, 2.0418).. controls (2.0026, 2.0675) and (2.0712, 2.0804) .. (2.1471, 2.0804).. controls (2.223, 2.0804) and (2.2916, 2.0675) .. (2.3529, 2.0418).. controls (2.4143, 2.0161) and (2.467, 1.9799) .. (2.5112, 1.9336).. controls (2.5551, 1.8873) and (2.5889, 1.8322) .. (2.6125, 1.7687).. controls (2.636, 1.705) and (2.6477, 1.6359) .. (2.6477, 1.5616).. controls (2.6477, 1.4856) and (2.636, 1.4163) .. (2.6125, 1.3533) -- cycle(3.8015, 1.7777).. controls (3.7825, 1.7534) and (3.7587, 1.7357) .. (3.7298, 1.7243).. controls (3.701, 1.7129) and (3.6687, 1.706) .. (3.6333, 1.7039).. controls (3.5975, 1.7015) and (3.5639, 1.7005) .. (3.5319, 1.7005) -- (3.3226, 1.7005) -- (3.3226, 2.0667) -- (3.5547, 2.0667).. controls (3.5864, 2.0667) and (3.6192, 2.0643) .. (3.6526, 2.0598).. controls (3.6859, 2.0553) and (3.7155, 2.0466) .. (3.7412, 2.0336).. controls (3.7669, 2.0206) and (3.7883, 2.0021) .. (3.805, 1.9778).. controls (3.8216, 1.9534) and (3.8301, 1.9209) .. (3.8301, 1.8799).. controls (3.8301, 1.8362) and (3.8206, 1.8021) .. (3.8015, 1.7777) -- cycle;
  \end{scope}
\end{tikzpicture}
}

\usepackage[affil-it]{authblk}
\usepackage{orcidlink}
\author[1%
  \ensuremath\mathparagraph]{Charles Tapley Hoyt%
    \,\orcidlink{0000-0003-4423-4370}\,%
    }

\affil[1]{RWTH Aachen University, Institute of Inorganic Chemistry%
    \,\protect\href{https://ror.org/04xfq0f34}{\protect\rorlogo}\,%
  }
\affil[$\mathparagraph$]{Corresponding author: %
  cthoyt@gmail.com %
}
\date{7 July 2025}

\begin{document}
\maketitle

\section{Statement of need}\label{statement-of-need}

Many modern cheminformatics workflows derive datasets from ChEMBL
(\citeproc{ref-Gaulton2017}{Gaulton et al., 2017};
\citeproc{ref-Zdrazil2023}{Zdrazil et al., 2023}), but few of these
datasets are published with accompanying code for their generation.
Consequently, their methodologies (e.g., selection, filtering,
aggregation) are opaque, reproduction is difficult, and interpretation
of results therefore lacks important context. Further, such static
datasets quickly become out-of-date. For example, the current version of
ChEMBL is v35 (as of December 2024), but ExCAPE-DB
(\citeproc{ref-Sun2017}{Sun et al., 2017}) uses v20, Deep Confidence
(\citeproc{ref-Cortes-Ciriano2019}{Cortés-Ciriano \& Bender, 2019}) uses
v23, the consensus dataset from Isigkeit et al.
(\citeproc{ref-Isigkeit2022}{2022}) uses v28, and Papyrus
(\citeproc{ref-Buxe9quignon2023}{Béquignon et al., 2023}) uses v30.
Therefore, there is a need for tools that provide reproducible bulk
access to the latest (or a given) version of ChEMBL in order to enable
researchers to make their derived datasets more transparent, updatable,
and trustworthy.

\section{State of the field}\label{state-of-the-field}

ChEMBL is typically accessed through its
\href{https://www.ebi.ac.uk/chembl/api/data/docs}{application
programming interface (API)}, through its Python client
(\citeproc{ref-Davies2015}{Davies et al., 2015}), through its RDF
platform (\citeproc{ref-Jupp2014}{Jupp et al., 2014}), or in bulk
through its
\href{https://ftp.ebi.ac.uk/pub/databases/chembl/ChEMBLdb/releases}{file
transfer protocol (FTP) server}. However, APIs and their respective
wrapper libraries are not efficient for querying and processing data in
bulk and bulk access is currently cumbersome for most potential users
due to the need to download, set up, and connect to databases locally.
Additional third-party software (e.g., KNIME, Pipeline Pilot) provide
alternate access to ChEMBL, but are relatively inflexible, inextensible,
or proprietary.

\section{Summary}\label{summary}

This article introduces \texttt{chembl-downloader}, a Python package for
the reproducible acquisition, access, and manipulation of ChEMBL data
through its FTP server.

At a low-level, it uses a combination of the
\href{https://github.com/cthoyt/pystow}{\texttt{pystow}} Python software
package and custom logic for the reproducible acquisition and
pre-processing (e.g., uncompressing) of either the latest or a given
version of most resources in the ChEMBL FTP server. These include
relational database dumps (e.g., in SQLite), molecule lists (e.g., in
SDF, TSV), pre-computed molecular fingerprints (e.g., in binary), a
monomer library (e.g., in XML), and UniProt target mappings (e.g., in
TSV).

At a mid-level, it provides utilities for accessing these files through
useful data structures and functions such as querying the SQLite
database with combination of Python's
\href{https://docs.python.org/3/library/sqlite3.html}{\texttt{sqlite3}}
library and \texttt{pandas} (\citeproc{ref-Mckinney2010}{McKinney,
2010}), parsing SDF files with RDKit (\citeproc{ref-rdkit}{Landrum,
n.d.}), parsing TSVs with \texttt{pandas}, loading fingerprints with
\texttt{chemfp} (\citeproc{ref-Dalke2019}{Dalke, 2019}), and parsing the
monomer library with Python's \texttt{xml} library. The low- and
mid-level utilities are kept small and simple such that they can be
arbitrarily extended by users.

At a high-level, it maintains a small number of task-specific utilities.
It contains several pre-formatted SQL queries to retrieve the
bioactivities associated with a given assay or target, to retrieve the
compounds mentioned in a publication or patent, to retrieve the names of
all compounds, etc.

\section{Case studies}\label{case-studies}

A first case study demonstrates how the high-level utilities in
\texttt{chembl-downloader} can be used to reproduce the dataset
generation from Cortés-Ciriano \& Bender
(\citeproc{ref-Cortes-Ciriano2019}{2019}), highlight some of the
methodological controversies (e.g., using arithmetic mean instead of
geometric mean for pIC\textsubscript{50} values), show the immense
variability introduced by new datapoints in later versions of ChEMBL,
and highlight the number of additional compounds added to each of its 24
target-specific datasets. Landrum \& Riniker
(\citeproc{ref-Landrum2024}{2024}) further investigated the impact of
such aggregation. See the corresponding
\href{https://github.com/cthoyt/chembl-downloader/blob/main/notebooks/cortes-ciriano-refresh.ipynb}{Jupyter
notebook}.

A second case study demonstrates the value of having a reproducible
script for identifying missing identifier mappings between molecules in
ChEMBL and ChEBI (\citeproc{ref-Hastings2016}{Hastings et al., 2016})
\emph{via} lexical mappings produced by Gilda
(\citeproc{ref-Gyori2022}{Gyori et al., 2022}), which identified 4,266
potential mappings for curation e.g., in a workflow such as Biomappings
(\citeproc{ref-Hoyt2022}{Hoyt et al., 2023}). See the corresponding
\href{https://github.com/cthoyt/chembl-downloader/blob/main/notebooks/chebi-mappings.ipynb}{Jupyter
notebook}.

A final case study demonstrates the utility of
\texttt{chembl-downloader} by making pull requests to the code
repositories corresponding to three popular cheminformatics blogs
(\href{https://greglandrum.github.io/rdkit-blog/}{the RDKit Blog},
\href{https://practicalcheminformatics.blogspot.com}{Practical
Cheminformatics}, and \href{https://iwatobipen.wordpress.com/}{Is Life
Worth Living?}) to make the code more reproducible (where the source
data was not available) and ChEMBL-version-agnostic:

\begin{itemize}
\tightlist
\item
  \href{https://github.com/greglandrum/rdkit_blog/pull/5}{greglandrum/rdkit\_blog
  (\#5)} for generating a substructure library
\item
  \href{https://github.com/PatWalters/sfi/pull/11}{PatWalters/sfi
  (\#11)} for calculating compounds' solubility forecast index,
  originally proposed by Hill \& Young (\citeproc{ref-Hill2010}{2010})
\item
  \href{https://github.com/PatWalters/jcamd_model_comparison/pull/1}{PatWalters/jcamd\_model\_comparison
  (\#1)} for comparing classification models
\item
  \href{https://github.com/iwatobipen/playground/pull/4}{iwatobipen/playground
  (\#4)} for parsing and using ChEMBL's monomer library
\item
  \href{https://github.com/iwatobipen/playground/pull/5}{iwatobipen/playground
  (\#5)} for analyzing chemical space of molecules from a set of patents
\end{itemize}

Additional external use cases can be found by
\href{https://github.com/search?q=\%22import\%20chembl_downloader\%22\%20OR\%20\%22from\%20chembl_downloader\%20import\%22\%20language\%3APython\%20NOT\%20is\%3Afork\%20-owner\%3Acthoyt&type=code}{searching
GitHub}. Finally, several scholarly articles, including from the ChEMBL
group itself, have used \texttt{chembl-downloader} in their associated
code (\citeproc{ref-Domingo-Fernuxe1ndez2023}{Domingo-Fernández et al.,
2023}; \citeproc{ref-Gadiya2023}{Gadiya et al., 2023};
\citeproc{ref-Gorostiola2024}{Gorostiola González et al., 2024};
\citeproc{ref-Nisonoff2023}{Nisonoff et al., 2023};
\citeproc{ref-Schoenmaker2025}{Schoenmaker et al., 2025};
\citeproc{ref-Zdrazil2023}{Zdrazil et al., 2023};
\citeproc{ref-Zhang2024}{Zhang et al., 2024}).

\section{Availability and usage}\label{availability-and-usage}

\texttt{chembl-downloader} is available as a package on
\href{https://pypi.org/project/chembl-downloader}{PyPI} with the source
code available at \url{https://github.com/cthoyt/chembl-downloader}
archived to Zenodo at
\href{https://doi.org/10.5281/zenodo.5140463}{doi:10.5281/zenodo.5140463}
and documentation available at
\url{https://chembl-downloader.readthedocs.io}. The repository also
contains an interactive Jupyter notebook tutorial and notebooks for the
case studies described above.

\section{Acknowledgements}\label{acknowledgements}

The author would like to thank Yojana Gadiya and Jennifer HY Lin for
helpful discussions and the NFDI4Chem Consortium
(https://www.nfdi4chem.de) for support.

\section*{References}\label{references}
\addcontentsline{toc}{section}{References}

\phantomsection\label{refs}

\end{document}